\documentstyle[12pt]{article}
\begin{document}

\font\fortssbx=cmssbx10 scaled \magstep2
\hbox to \hsize{
%\special{psfile=/NextLibrary/TeX/tex/inputs/uwlogo.ps
%			      hscale=8000 vscale=8000
%			       hoffset=-12 voffset=-2}
\hskip.5in \raise.1in\hbox{\fortssbx University of Wisconsin - Madison}
\hfill$\vcenter{\hbox{\bf MADPH-95-901}
          \hbox{July 1995}}$ }

\vspace{.5in}

\begin{center}
{\small\bf GAMMA RAY ASTRONOMY WITH UNDERGROUND DETECTORS}\\[.1cm]
{\tenrm F.~HALZEN}\\
\tenit Department of Physics, University of Wisconsin, Madison, WI 53706,
USA\\[.1cm]
{\tenrm T.~STANEV}\\
{\tenit Bartol Research Institute, University of Delaware, Newark, Delaware
19716, USA}
\end{center}

\bigskip

{\small
\begin{center}ABSTRACT \end{center}
{\narrower
Underground detectors measure the directions of up-coming muons of
neutrino origin. They can also observe down-going muons made by gamma rays
in the Earth's atmosphere. Although gamma ray showers are muon-poor,
they produce a sufficient number of muons to detect the sources
observed by GeV and TeV telescopes. With a threshold higher
by one hundred and a probability of muon production of about
$1\%$ for the shallower AMANDA and Lake Baikal detectors,
these instruments can, for a typical GRO source, match the
detection efficiency of a GeV satellite detector since their effective
area is larger by a factor $10^4$. The muons must have enough energy
for accurate reconstruction of their direction. Very energetic muons on the
other hand are rare because they are only produced by higher energy
gamma rays whose flux is suppressed by the decreasing flux at the
source and by absorption on interstellar light.  We show that there
is a window of opportunity for muon astronomy in the 100~GeV energy
region which nicely matches the threshold energies of the AMANDA
and Lake Baikal detectors.
\par}}

\newpage
\section{Introduction}

Instruments exploiting the air Cherenkov technique have extended the
exciting astronomy revealed by the Compton G(amma) R(ay) O(bservatory)
into the TeV energy range\cite{Weekes}. The photon spectra of the Crab
nebula, the pulsar PSR 1706-44 and of the active galaxy Markarian 421,
observed by satellite experiments at GeV energies, are known to extend
to higher energy, e.g.\ all the way to 10~TeV for the Crab. More
interestingly, the galaxy Markarian 501, recently detected in the TeV
range, is not a confirmed GeV gamma ray source.

TeV gamma rays produce muons in the Earth's atmosphere that can be detected
and reconstructed in relatively shallow underground detectors such as the
AMANDA and Lake Baikal detectors, which are positioned at a
modest depth of order 1~kilometer~\cite{GHS}. They are therefore sensitive
to muon energies of a few hundreds of GeV, well below the TeV thresholds
of deep underground detectors. Although muons from such sources compete
with a large background of down-going cosmic ray muons, they can be
searched for with relatively large effective area detectors. Unlike air
Cherenkov telescopes, muon detectors cover a large fraction of the sky.
Useful results may, possibly, be obtained with the partially deployed
instruments, even before they achieve the necessary up-down discrimination
to identify neutrinos. Moreover,
background multi-muon bundles, which are difficult to reconstruct, can be
conveniently rejected without suppression of the predominantly single-muon
gamma ray signal.

In this paper we demonstrate how the 100~GeV muon energy range provides
a window of opportunity for muon astronomy. The muons are sufficiently
energetic to leave tracks that can be adequately measured by the Cherenkov
technique. The direction of the parent photon can be inferred with
degree accuracy. Hundred-GeV muons originate in TeV gamma showers
whose existence has been demonstrated, at least for two galactic
and two extra-galactic sources, by air-Cherenkov telescopes.
A multi-TeV air shower will produce a 100~GeV muon with a
probability of order 1\%\cite{HHS}, sufficient to observe the
brightest sources with relatively modest size detectors with
effective area of order 1000~m$^2$ or more. The probability
that such photons produce TeV-energy muons which trigger the deep underground
detectors, such as those in
the Gran Sasso tunnel, is small. TeV muons are produced by photons
of several tens of TeV energy and above. The rates are however
suppressed and, more importantly, it is not clear whether the spectra
extend far beyond the TeV region. They most likely do not. In the
case of galactic sources, such as the Crab supernova remnant,
the current thinking is that the high energy photons are produced
by inverse Compton scattering of electrons accelerated by the
pulsar\cite{HdeJ}. Such a purely electromagnetic accelerator is
unlikely to produce photons far beyond the observed spectrum which
extends to 10~TeV. While the vast majority of GeV gamma ray sources
display a $E^{- \gamma}$ energy spectrum with $\gamma
\simeq 1$\cite{2EC}, the steepening slope of the Crab spectrum may
provide us with a glimpse of a steep cutoff not far beyond the reach
of the data; see Table~1. On the contrary, extra-galactic sources such
as the active galaxy Markarian 421, may be true
high energy accelerators producing protons with sufficient energy
to account for the high energy cosmic ray spectrum which extends
beyond $10^{20}$~eV. This does not guarantee emission of
high energy photons which may be absorbed in the source, or
on the interstellar infrared and microwave background\cite{SS}.
Very near active galactic nuclei, in the local cluster or the
super-galactic plane and with redshift less than 0.03 or so,
represent promising sources in this respect. Examples are listed in
Table~1 where, as usual, we have parametrized the gamma ray flux as
\begin{equation}
{dN_\gamma\over dE_\gamma} = {F_\gamma   \over E^{(\gamma+1)}} 10^{-12} \rm\,
cm^{-2} \, s^{-1}\;.
\end{equation}

Throughout this paper energies are in TeV units. The high energy
luminosity of the source is described by the parameter $F_{\gamma}$
which in the EGRET catalog\cite{2EC} denotes the flux of photons
above 100~MeV in units of $10^{-8}$ photons per cm$^2$ per second.
For flat ($\gamma = 1$) spectra the same number will apply to the
flux of TeV gamma rays in units $10^{-12}$. Most EGRET sources with
measured energy spectrum have $\gamma \leq 1$. For the Crab supernova
however the TeV flux is reduced by one order of magnitude with
$F_{\gamma} = 20$ in the TeV region\cite{WhippleC}.
This number is bracketed by the 7--50 Markarian 421 flux in its low and
high states\cite{WhippleM}. So, interestingly, galactic and nearby
extra-galactic sources produce comparable
photon fluxes at Earth despite the $10^5$ ratio of their distances.
Sources such as the Vela pulsar, the galactic center and the cluster
of four unidentified gamma ray sources in the direction of the spiral
arm near Cygnus, may be TeV gamma ray emitters brighter by more than
one order of magnitude. We refer to Table~1 for a partial list.

\bigskip
\begin{table}[h]
\centering
\tabcolsep.75em
\caption{A partial list of potential VHE $\gamma$-ray sources based on the
2nd EGRET catalog. Groups of sources that are difficult to resolve are
combined. The position for such sources are averaged with the EGRET
$\gamma$-ray flux and a solid angle (ster) for the group is given.
$F_\gamma$ is the average number of photons above 100 MeV in units
of $10^{-8}\rm\,cm^{-2}\,s^{-1}$.}
\smallskip
\begin{tabular}{rrr@{$\,\pm\,$}lcccc}
\hline
RA& Dec& \multicolumn{2}{c}{$F_\gamma$}& $\gamma_d$& $\Delta\Omega$&
Source& VHE? \\
\hline
128.8& $-$45.2& 932& 8& 1.6--1.9& ---& Vela pulsar& $\surd$\\
98.5& 17.8& 374& 5& 1.4--1.7& ---& Geminga pulsar& $\surd$\\[2mm]
306.1& 39.2& 335& 9& ---& $3.3\times10^{-3}$&
$\left\{\parbox[c]{.875in}{J2019+3719\\ J2020+4026\\ J2026+3610\\
J2033+4122}\right.$& ---\\[14mm]
266.8& $-$29.8& 218& 10& ---& $2.7\times10^{-4}$&
$\left\{\parbox[c]{.875in}{J1746$-$2852\\ J1747$-$3039}\right.$& ---\\[2mm]
184.56& $-$5.8& 212& 3& 2.2--2.5& ---& Crab pulsar& $\surd$\\
257.4& $-$44.9& 144& 5& ---& ---& PSR1706$-$44& $\surd$\\[2mm]
217.7& $-$23.4& 121& 9& ---& $3.6\times10^{-4}$&
$\left\{\parbox[c]{.875in}{J1801$-$2312\\ J1811$-$2339}\right.$& ---\\[2mm]
213.2& $-$62.2& 103& 9& ---& ---& J1412$-$6211& ---\\
155.4& $-$58.6& 99& 5& ---& ---& J1021$-$5835& ---\\
\hline
\end{tabular}
\end{table}

\section{Muon Rates from Gamma Ray Sources}

Gamma rays initiate atmospheric cascades of not only electrons and photons,
but also muons. Muons originate from the decay of pions which are
photoproduced by shower photons. The number of muons with energy
above $E_{\mu}$ in a shower initiated by a photon of energy $E_{\gamma}$
has been computed some time ago and for $E_{\mu}$ in the range from
100 to 1000 GeV can be parametrized as
\begin{equation}
N_\mu (E_\gamma, > E_\mu) \cong {2.14\times 10^{-5} \over \cos\theta}
{1\over (E_\mu/\cos\theta)} \left[ E_\gamma \over (E_\mu/\cos\theta) \right]
\end{equation}
for $E_{\gamma}/E_{\mu} \geq 10$. This estimate is conservative and below
the rate of muons obtained by Bhattacharyya\cite{Bhat}, who updated the
calculations of reference\cite{HHS} taking into account the latest
measurements of the high energy photoproduction cross section at HERA.
Additional TeV-muons are produced by pair production and the decay of
charm particles. For lower energy muons these additional sources can
be neglected\cite{HHS}. The muon flux produced by a gamma ray source
is obtained by convolution of Eqs.~1 and~2:
\begin{eqnarray}
N_\mu ({>}E_\mu) &=& \int_{E_{\gamma\,{\rm min}}}^{E_{\gamma\,{\rm max}}}
dE_\gamma
{F_\gamma 10^{-12} \over E_\gamma^{\gamma+1}}\,{2.14\times 10^{-5} \over E_\mu}
\left( E_\gamma \cos\theta \over E_\mu \right) \\
&\simeq& 2\times 10^{-17} {F_\gamma\over \cos\theta}
{1\over (E_\mu/\cos\theta)^{\gamma+1}}\, \ln\!\!\left(E_{\gamma\,{\rm max}}
\over E_{\gamma\,{\rm min}} \right) f \;.
\end{eqnarray}
Here $E_{\mu}$ is the vertical threshold energy of the detector,
e.g.\ 0.18~TeV for the AMANDA detector. Photons with energy ranging from
a minimum energy $E_{\gamma\,{\rm min}} \simeq 10 \times E_{\mu} / cos\theta$
to the maximum energy of the source $E_{\gamma\,{\rm max}}$ mainly
contribute to the production of the observed muons. The highest energy
photons dominate. For this reason the muon flux critically depends on
the high energy flux of the source. $\theta$ is the zenith angle at which
the source is observed. This angle is, conveniently, time-independent
for the AMANDA detector with a South Pole location. The factor $f$
is a correction factor which can be parametrized as
\begin{equation}
f = \left(E_\mu/\cos\theta\over 0.04\right)^{0.53} \;.
\end{equation}
The flux of muons varies with vertical threshold as $E_{\mu}^{-(\gamma+1)}$.
This behavior is only approximate and assumes that the integrand in Eq.~3
spans many decades of the $E_{\gamma}^{-2}$ spectrum between
$E_{\gamma\,\rm min}$ and $E_{\gamma\,\rm max}$. Otherwise, the dependence
is moderated, an effect which is described by the factor $f$. In the end
our parametrization reproduces the explicit Monte Carlo results\cite{HHS}.

Above signal has to be extracted from a background of cosmic ray muons
which is empirically $2 \times 10^{-7}$ muons per cm$^2$ per second per
steradian and falls with zenith angle as $\cos\theta^{2.8}$ at a detector
depth of 1~kilometer\cite{Baikal}. Relevant is the number of background
muons in a pixel of $\delta \times \delta$ degrees which is given by
\begin{equation}
N_\mu^{\rm back} (\rm m^{-2} \, yr^{-1}) \simeq 20\cos\theta^{2.8} \delta^2 \;.
\end{equation}

As previously mentioned the background includes some fraction of
multi-muon events. Rejecting multi-muon events not only improves
signal-to-noise, it should improve angular resolution which is often
degraded by the poor reconstruction of complex muon bundles initiated
by high energy cosmic ray muons.

\section{Summary of Results and Examples}

Our results can be conveniently summarized as follows. The number of
events per year in a detector of effective area $A$ m$^2$ is given by
\begin{equation}
N_\mu ({\rm yr^{-1}}) = 6.7\times 10^{-5} {F_\gamma\over\cos\theta}
{1\over(E_\mu/\cos\theta)^{\gamma+1}} \,
\ln\!\!\left(\cos\theta E_{\gamma\,{\rm max}} \over 10E_\mu\right) f A \;.
\end{equation}
We recall that all energies are in TeV units. The signal-to-noise,
defined as the number of events divided by the square root of the
number of background events, in a pixel of $\delta \times \delta$ degrees
depends on detector area and zenith angle as
\begin{equation}
S\Big/\sqrt N \sim {\sqrt A\over \cos\theta^{0.9} \delta} \;.
\end{equation}
The formula simply expresses that signal-to-noise is improved for
increased area, better resolution and for sources at large zenith
angle in the sky where the cosmic ray background muon rate is reduced.

To demonstrate the power of a neutrino telescope we start with an
optimistic, though not unrealistic example. We take the Vela pulsar
with $F_{\gamma} = 932$ and $\theta = 45^\circ$ at the South Pole.
This assumes that the source flux is not cut off at high energy. We will
however assume that $E_{\gamma\,\rm max}$ is only 10~TeV (the muon flux is
increased by a factor 4.4 if the spectrum extends to 1000~TeV). For a
nominal detector with effective area $10^4$~m$^2$ area and $\delta = 1^\circ$
angular resolution we obtain 5000~events per year on a background
of $7.5 \times 10^4$ or a signal to noise in excess of 10. $S/\sqrt N$
exceeds unity for a detector as small as 40m$^2$ collecting 20~events
per year. For a detector with a poor 5$^\circ$ resolution this area
is 200m$^2$. The partially deployed AMANDA and Lake Baikal instruments
may already be sensitive to such a source.

For the blazar Markarian~421 the TeV-flux, averaged between the high
and low state, corresponds to $F_{\gamma} \simeq 35$. Our nominal
detector should collect $1.8\times10^2 (8.2\times10^2$) events
per year assuming $E_{\gamma} =10$(100~TeV) for a $S/\sqrt N$ of 0.7(2.8).
For a blazar known to emit 10~TeV gamma rays the assumption of a
cut-off of only 10~TeV is likely to be conservative.

As a final example we propose the 4 sources in a 5$^\circ$ by 3$^\circ$
declination/right ascension bin in the direction of the spiral arm in
Cygnus. For this cluster of sources $F_{\gamma} = 335$ and
$\theta = 60^\circ$; we will assume that $E_{\gamma} =
 100$~TeV. A $10^4\rm~m^2$ detector will collect 6300~events on a
background of 0.5~million in the bin containing the sources for
$S/\sqrt N = 8.9$. No precise reconstruction is required and the
sources should be detected, even if their spectrum cuts off at 10~TeV.

\section*{Acknowledgments}

This research was supported in part by the U.S.~Department of
Energy under Grant No.~DE-FG02-95ER40896 and in part by the
University of Wisconsin Research Committee with funds granted by
the Wisconsin Alumni Research Foundation.

\end{document}